\documentclass[prl,twocolumn,nofootinbib]{revtex4}

\usepackage{graphicx}
\usepackage{color}
\usepackage{dcolumn}

\def\beq{\begin{eqnarray}}
\def\eeq{\end{eqnarray}}

\begin{document}

\title{A new approach to resummation: Parametric Perturbation Theory}
\author{Paolo Amore}\email{paolo@ucol.mx} 
\affiliation{Facultad de Ciencias, Universidad de Colima, \\
Bernal D\'{\i}az del Castillo 340, Colima, Colima, Mexico} 

\begin{abstract}    
We present a {\sl non--perturbative} method, called {\sl Parametric Perturbation Theory} (PPT), which is
alternative to the ordinary perturbation theory. The method relies on a principle of simplicity for 
the observable solutions, which are constrained to be linear in a certain (unphysical) parameter. The perturbative
expansion is carried out in this parameter and not in the physical coupling (as in ordinary perturbation theory).
We show that the method is capable to resum the divergent perturbative series, to {\sl extract} the leading asymptotic 
(strong coupling) behavior and {\sl predict} with high accuracy the coefficients of the perturbative series. 
\end{abstract}
\maketitle


In this letter we present a method, called Parametric Perturbation Theory (PPT), 
which can be used to resum series, either divergent or with a finite radius
of convergence, which appear in the perturbative solution of physical problems. The approach behind our method 
is completely new and it is based on few simple ideas: the first idea, which we call {\sl Principle of Absolute Simplicity} (PAS), 
is that, instead of expanding the observable (energy, frequency, $\dots$) in the physical coupling $g$, and thus obtain 
to a finite order a polynomial in $g$, we impose that the observable has the simplest possible form (linear) in a given 
unphysical parameter $\varrho$; the second idea is that the expansion must be carried out in $\varrho$ and that the 
functional relation $g = g(\varrho)$ (unknown) must comply with the PAS to the order to which the calculation is done. 
This will allow to determine the relation between $g$ and $\rho$ and in turn to obtain the observable as a parametric function
of $\varrho$. 

Consider for example a model which depends on a parameter $g$, and which is solvable when $g=0$. 
Clearly, the application of Perturbation Theory (PT) to the calculation of a physical observable $\mathcal{O}$ 
to a finite order yields a polynomial in $g$. Calling $r$ the radius of convergence of the perturbative series, 
the direct use of PT must be restricted to $|g|<r$. However, the misbehavior of the perturbative series 
for $\mathcal{O}$ is the result of having expanded in a parameter, $g$, which is not optimal. 
As a matter of fact, if one knew the exact solution to the problem, i.e. $\mathcal{O} = f(g)$, then  
this solution could be considered as a polynomial of order one in the variable $\varrho = f(g)$. 
Although this observation by itself cannot be used as a constructive principle, we may adopt
the philosophy that the {\sl perturbative series for the observable can be simpler and  convergent in all the 
domain, if it is cast in terms of a suitable parameter $\varrho$}. 
Only if such parameter, by luck or ability, turns out to be the $\varrho = f(g)$ discussed above, 
the exact solution is obtained. The goal, therefore, is to progressively build this parameter $\varrho$ 
to yield an expression for $\mathcal{O}$ as simple as possible. 
In this framework the perturbative expansion is carried out in $\varrho$ and all the physical quantities 
in the problem are expressed as functions of $\varrho$. In particular we have now that $g = g(\varrho)$. 
While the ordinary perturbation theory works by calculating the contributions to higher orders in $g$, 
each term of higher order refining the result to lower order, the approach approach is the opposite: we carry 
out a perturbative  calculation in $\varrho$, and then determine order by order the form of $g = g(\varrho)$ so 
that the observable $\mathcal{O}(\varrho)$ can be a order one polynomial in $\varrho$, as required by 
the {\sl Principle of Absolute Simplicity}.

Although in this letter we focus on the implementation of this method as a technique for resumming 
perturbative series, we show in a companion and more detailed paper that the same philosophy can be 
used to obtain a perturbation scheme fully autonomous from perturbation theory. 

Let us first sketch briefly how the method works and then apply it to some non-trivial perturbative series.
Consider an observable  $\mathcal{O}$, represented through the perturbative series  perturbatively to some order
\beq
\mathcal{O}^{(PT)} = \sum_{n=0}^{N} b_n \ g^n \ ,
\label{eq_1}
\eeq
where $N=\infty$ in some cases. The series could be either convergent, with a finite radius of convergence,
or divergent, although we will not worry at this time. The full implementation of the method requires the introduction
of an unphysical parameter, $\varrho$, and the specification of a functional relation between $g$ and $\varrho$:
\beq
g = g(\varrho) \ .
\label{eq_2}
\eeq

The choice of this relation is not completely arbitrary, since it must be capable of reproducing the perturbative
terms when expanded around $g=0$. For example, in many of the cases which we have studied we have used
\beq
g(\varrho) = \varrho \ \left[ \frac{1+\sum_{n=1}^{\bar{N}_u} c_n \varrho^{n}}{1+\sum_{n=1}^{\bar{N}_d} 
d_n \varrho^{n}}  \right] ,
\label{eq_3}
\eeq
where the coefficients $c_n$ and $d_n$ are unknown to be determined applying the PAS. $\bar{N}_u$ and $\bar{N}_d$  
are integers.

The {\sl Principle of Absolute Simplicity} requires that the observable be linear in $\varrho$, i.e.
\beq
\mathcal{O}^{(PPT)} = b_0 + b_1 \varrho \ .
\label{eq_4}
\eeq

If we substitute eq.~(\ref{eq_3}) inside eq.~(\ref{eq_1}) and expand around $\varrho = 0$, working to a given order
specified by the sum $\bar{N}_u+\bar{N}_d=\bar{N}$, then we can fully determine the coefficients $c_n$ and $d_n$
by requiring that an equal number of nonlinear terms in $\varrho$ vanish, starting with the term going as $\varrho^2$.
Notice that the choice of the integer parameters  $\bar{N}_u$ and $\bar{N}_d$ determines the asymptotic behavior
of $g$ as $\varrho \rightarrow \infty$ (we are assuming for simplicity that the denominator in (\ref{eq_3}) never vanish
for $\varrho>0$):
\beq
g(\varrho) \approx \varrho^{\bar{N}_u-\bar{N}_d+1}
\label{eq_5}
\eeq
and in turn
\beq
\mathcal{O}^{(PPT)} \approx b_0 + b_1 g^{1/(\bar{N}_u-\bar{N}_d+1)} \ .
\label{eq_6}
\eeq

Therefore in cases where the asymptotic behavior is known, one can choose  $\bar{N}_u$ and $\bar{N}_d$  to reproduce the exact 
asymptotic behavior of the solution; on the other hand, when the asymptotic behavior is unknown, working to a given order 
$\bar{N}_u+\bar{N}_d=\bar{N}$, one can select the most appropiate asympotic behavior among those allowed by the combination
of $\bar{N}_u$ and $\bar{N}_d$  which keep $\bar{N}$ fixed.

We will first apply the method to a model of $\phi^4$ field theory in zero dimensions~\cite{JZJ}, which will allow us to discuss
some other properties of our resummation. We consider the integral
\beq
E(g) = \int_{0}^{+\infty} e^{-x^2-g x^4} dx \ ,
\label{eq_7}
\eeq
whose perturbative series is divergent.  Eq.(\ref{eq_7}) admits an exact analytical solution given by
\beq
E(g) = \frac{e^{\frac{1}{8 g}}}{4 \sqrt{g}} \ K_{1/4}\left(\frac{1}{8 g}\right) \ ,
\label{eq_8}
\eeq
where $K_{1/4}(g)$ is the Bessel function of order $1/4$. Notice that for negative values 
of $g$ this expression acquires an imaginary part, signaling that the system becomes metastable. 

We will now analyze this problem with the help of PPT, using a slightly different functional relation
than the one in eq.~(\ref{eq_3}):
\beq
g(\varrho) = \varrho \ \left[ \frac{1+\sum_{n=1}^{\bar{N}} c_n \varrho^{n}}{1+\sum_{n=1}^{\bar{N}+1} 
d_n \varrho^{n}}  \right]^5 .
\label{eq_9}
\eeq
Notice that this relation implements the correct asymptotic behavior, $E(g) \propto g^{-1/4}$ as 
$g\rightarrow \infty$.

Using $\bar{N}=3$ we have found the transformation:
\beq
g(\varrho) \approx \frac{\rho  \left(2925 \varrho^3+ 881.8 \varrho^2+ 58.8 \varrho +1\right)^5}{\left(
-1737.2 \varrho^4+ 2243.9 \varrho^3+ 832.2 \varrho^2+ 58\varrho +1\right)^5} \ .
\label{eq_10}
\eeq

\begin{figure}
\begin{center}
\bigskip\bigskip\bigskip
\includegraphics[width=7cm]{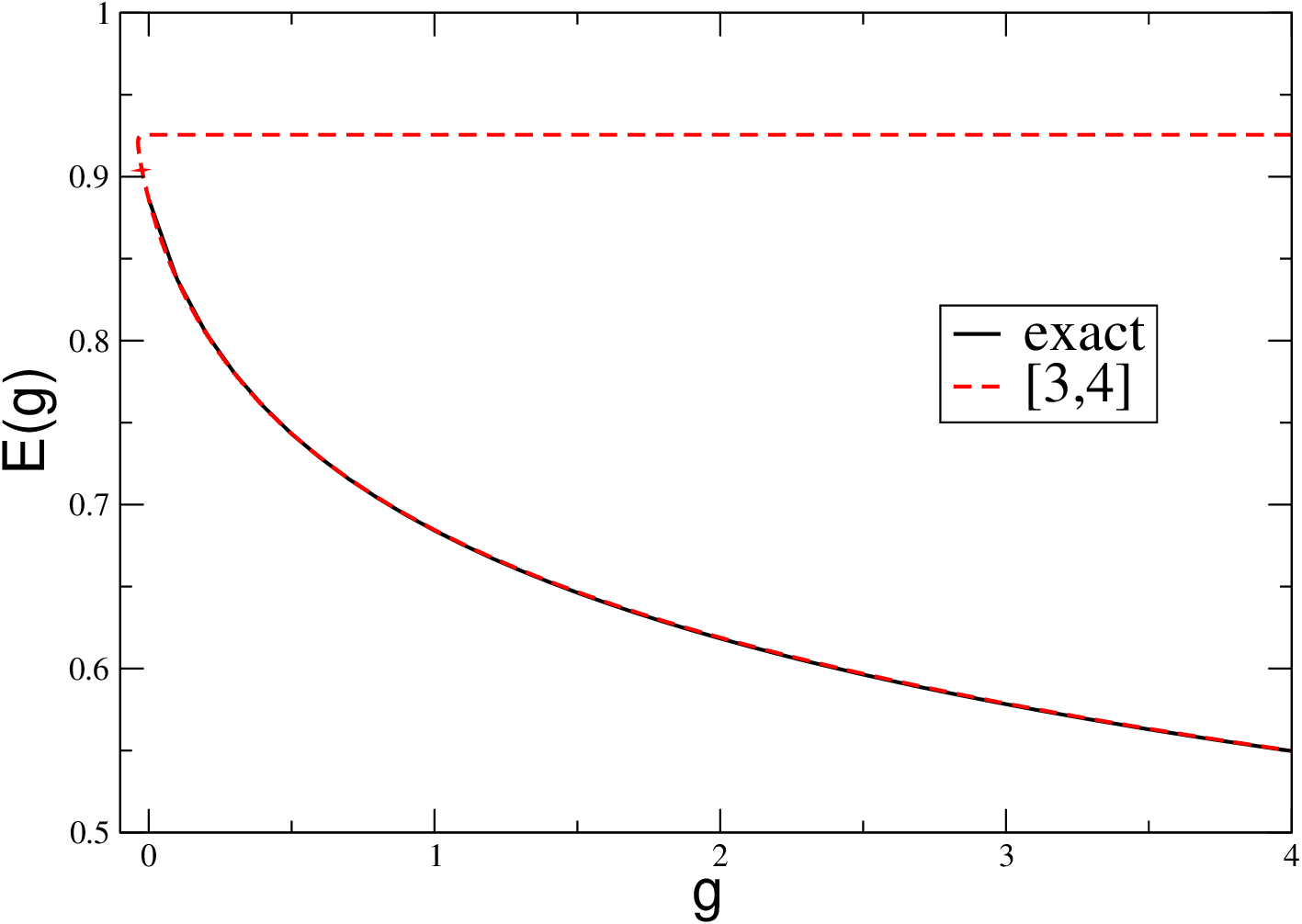}
\caption{(color online) 
Comparison between the set $[3,4]$ and the exact integral. The approximate solution has a 
branch point close to $g=0$.}
\label{Fig_1}
\end{center}
\end{figure}

Since $\varrho_0 = - 0.0593$ is a zero of the denominator, $\lim_{\varrho \rightarrow \varrho_0} g(\varrho) = \infty$:
this result signals the presence of a branch point in the proximity of $\varrho_0$ (see  Fig.\ref{Fig_1}).
The existence of this branch point can be understood as a signal of the no-analyticity of the exact solution
in $g=0$, which is the reason why the perturbative series diverges. 

In the region $g < 0$ the analytic continuation of the solution acquires an imaginary part.  
Using eq.(\ref{eq_10}) we find the numerical solutions 
of $g(\varrho) = g$, with $g<0$. For example, corresponding to $g=-1$ we obtain two pairs of complex conjugated 
roots accompanied by a single real root:
\beq
\varrho_1 &=& 0.20784+0.54897 \ i \\
\varrho_2 &=& 0.20784-0.54897 \ i \\
\varrho_3 &=& -0.24838 \\
\varrho_4 &=& -0.22321+0.00636 i \\
\varrho_5 &=& -0.22321-0.00636 i \ .
\eeq

Obtaining a complex value for $\varrho$ has an immediate effect on the observable $E(g)$, which 
{\sl acquires an imaginary part}, $Im E(g) = b_1 \ Im \varrho$.
Therefore we can verify if one of these solutions corresponds to the exact solution of (\ref{eq_7}) for $g=-1$:
\beq
Im E(g) = -0.37679
\eeq
which should be compared to the imaginary parts calculated with the PPT using the numerical roots $\varrho_i$, 
$i=1,\dots, 5$:
\beq
Im E(g)^{PPT}_1 &=& -0.36488 \\
Im E(g)^{PPT}_2 &=& 0.364886 \\
Im E(g)^{PPT}_3 &=& 0 \\
Im E(g)^{PPT}_4 &=& -0.00422 \\
Im E(g)^{PPT}_5 &=& 0.00422  \ .
\eeq

\begin{figure}
\begin{center}
\bigskip\bigskip\bigskip
\includegraphics[width=7cm]{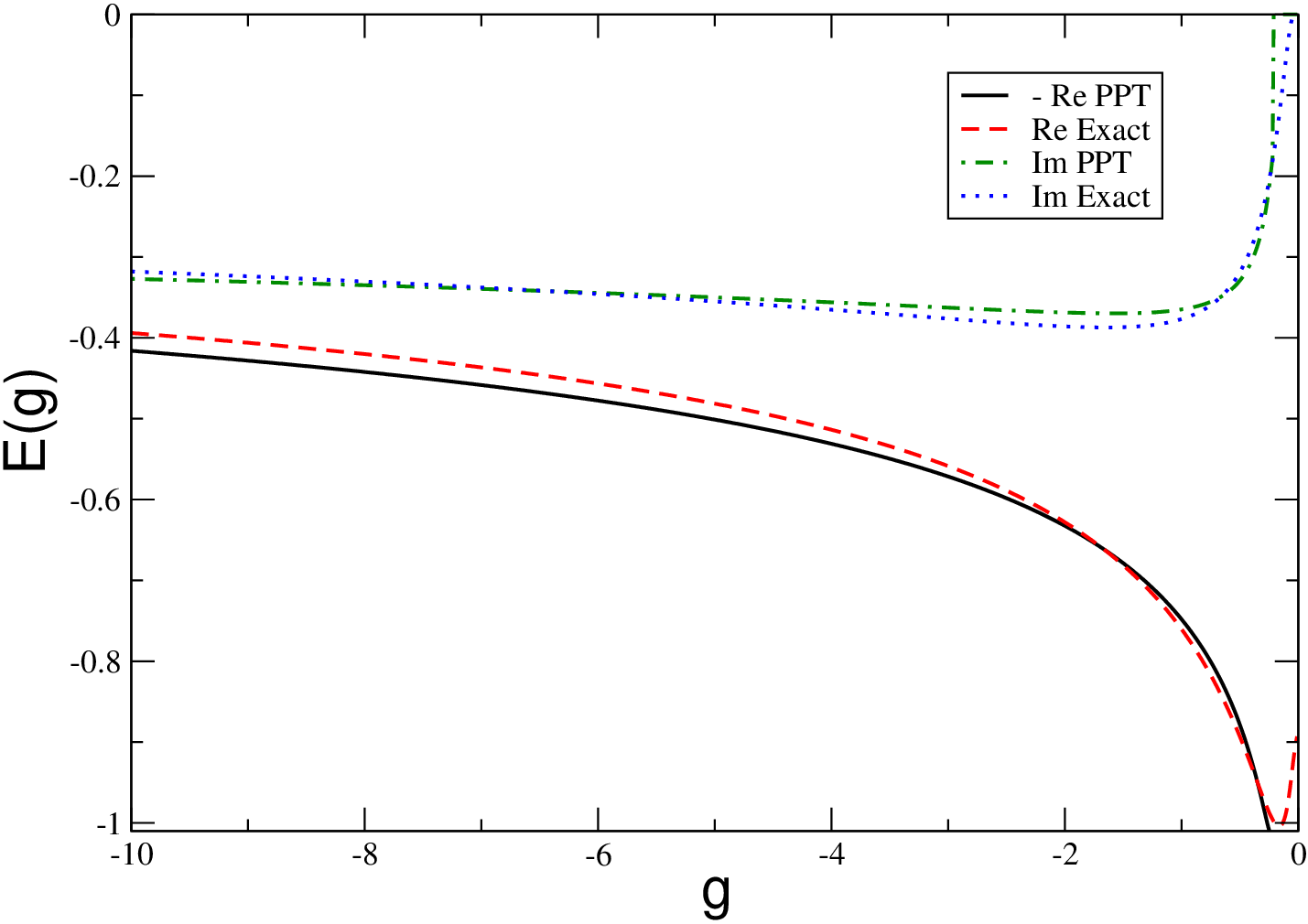}
\caption{(color online) 
Real and imaginary parts of $E(g)$ for the zero dimensional $\phi^4$ theory obtained using the PPT with $\bar{N} = 3$. 
The results are compared with the exact expression of eq.(\ref{eq_7}).}
\label{Fig_2}
\end{center}
\end{figure}

The comparison between exact and approximate real and imaginary parts of the integral for negative $g$ is shown
in Fig.\ref{Fig_2}.

These results suggest that the first root corresponds to the analytic continuation of the solution for $g>0$
to negative values. On the other hand the real part of $Re E(g)^{PPT}_1 = 0.74808$ has the opposite
sign of $Re  E(g) = -0.76033$: this happens because our function is continous and therefore 
it is not possible to reproduce a discontinuity at $g=0$.
The exponential  behavior of the exact solution for $g\rightarrow 0^-$ cannot  be reproduced 
in this approach.

We will now discuss a different problem, a lattice $\phi^4$ model in $1+1$ dimensions, described by
the hamiltonian~\cite{Nish01}
\beq
\hat{H} =  \sum_i \left[ \frac{1}{2} \hat{\pi}_i^2 + \hat{\phi}_i^4 + g \left( \frac{1}{2} 
\left(\hat{\phi}_i-\hat{\phi}_{i+1}\right)^2+ \frac{1}{2} \hat{\phi}_i^2 \right)\right] \ ,
\eeq
where $i$ is the site index and $\hat{\pi}_i$ and $\hat{\phi}_i$ are canonically conjugated operators.

Nishiyama has studied this model using both a linked cluster expansion, calculating the perturbative
contributions to order $g^{11}$, and  the Density Matrix Renormalization Group (DMRG).
Since the perturbative series has a radius of convergence $g_0 \approx 1$, he used an Aitken $\delta^2$ 
process to accelerate the convergence of this series, obtaining moderately improved results. Using these
results, he speculated the existence of a singularity corresponding to $g\approx -2$ (using our notation) and
of a Ising-type phase transition for a critical negative $g$.

We can apply PPT to this problem using eq.(\ref{eq_3}) and considering the sets corresponding to $\bar{N}=5$.
Comparing the difference $E(g)^{PT}-E(g)^{PPT}$ we have found that the optimal set corresponds to 
$[\bar{N}_u,\bar{N}_d]=[3,2]$. 
In Table \ref{table1} we compare the exact perturbative coefficients going from $b_7$ to $b_{11}$ 
as given in \cite{Nish01} with those {\sl predicted} by the set $[3,2]$.  The largest error corresponds 
to $b_{11}$ and is of just $0.1  \%$!

\begin{widetext}
\begin{table}
\caption{Comparison between the perturbative coefficients of \cite{Nish01} and those predicted with
PPT working with the set $[3,2]$.}
\begin{tabular}{cccccc}
\colrule
              & $b_7$  & $b_8$ & $b_9$ & $b_{10}$ &  $b_{11}$ \\
\colrule
$b_n^{exact}$ & 0.01106139 & -0.00874935 & 0.00709675 & -0.00587143 & 0.00493622 \\
$b_n^{[3,2]}$ & 0.01106113 & -0.00874838 & 0.00709460 & -0.00586767 & 0.00493037 \\
error ($\%$) & 0.00233 & 0.011081739435 & 0.03022 & 0.06403 & 0.11851 \\
\colrule
\end{tabular}
\label{table1}
\end{table}
\end{widetext}

\begin{figure}
\begin{center}
\bigskip\bigskip\bigskip
\includegraphics[width=7cm]{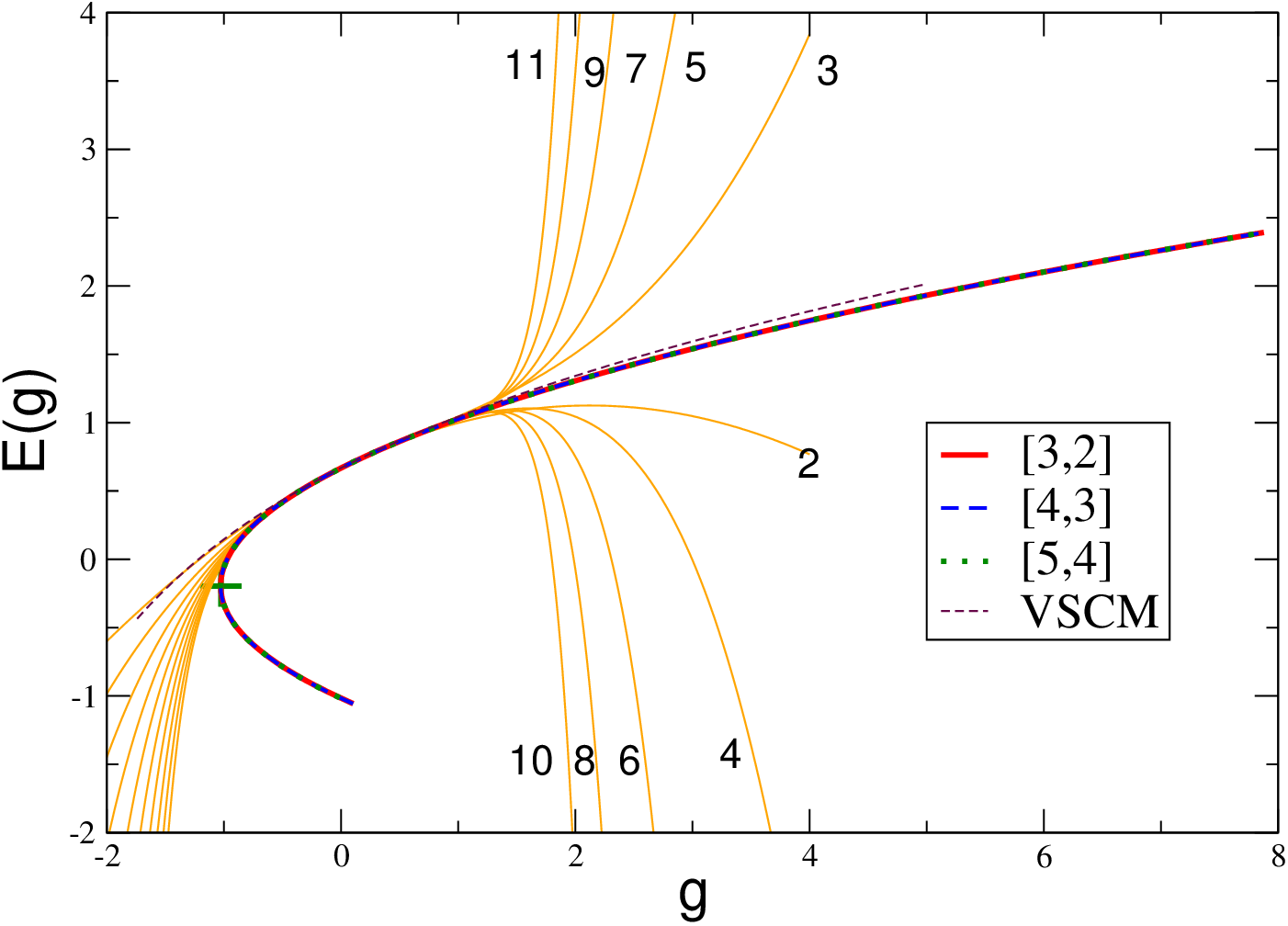}
\caption{(color online) 
Comparison between the resummed energies to orders $[3,2]$, $[4,3]$ and $[5,4]$ and the perturbative 
polynomials.}
\label{Fig_3}
\end{center}
\end{figure}

In Fig.\ref{Fig_3} we have compared the perturbative polynomials for the energy from orders $g^2$ to 
$g^{11}$ with the energy resummed with the sets $[3,2]$, $[4,3]$ and $[5,4]$. There are several striking 
aspects which should impress the reader: first of all, the difference between the three sets is extremely 
thin, thus signaling that the convergence is extremely strong; in second place, the resummed energy 
confirms the DMRG result displayed in Fig.2 of \cite{Nish01}; finally, the resummed energy is a multivalued 
function, with a branch point at $g \approx -1.025$, falling extremely close to the singularity speculated by
Nishiyama~\footnote{Clearly, the branch point of a function $y = f(x)$ at a point $x=x_0$ 
manifests itself as a singularity in that point when it is calculated using the Taylor series around a 
different point.}.
Because of the use of a parameter $\varrho$, PPT can deal with multivalued functions in a way which is not 
possible in conventional perturbation theory. 
Finally, the thiny dashed line in the plot corresponds to the numerical result obtained in \cite{Amore06a} using
the Variational Sinc Collocation Method (VSCM) within a mean field approach.

As a last example, we apply PPT to the prediction of virial coefficients of a hard sphere gas in $D=2,3$ dimensions.
Ref.\cite{Clis06} contains the values for the first $10$ virial coefficients for $D=2,\dots, 8$.
In this case, we have found out that the optimal choice corresponds to $\bar{N}_u = \bar{N}_d-1$, 
thus implying (using our notation) that  $\lim_{\varrho\rightarrow \infty} g(\varrho) = \bar{g} < \infty$,
i.e. that {\sl the resummed function will have a singularity precisely at $g=\bar{g}$}.

Working with the set $[3,4]$ we have found
\beq
\bar{g}^{[3,4]}_{D=2} \approx 1.1625 \ \ , \ \ \bar{g}^{[3,4]}_{D=3} \approx 1.43439 \ .
\eeq
 
In Table \ref{table2} we have considered the different sets corresponding to $\bar{N}_u +\bar{N}_d= 5$, 
which use the first eigth virial coefficients as input, and used them to predict $B_9$ and $B_{10}$, which
have already been calculated~\cite{Clis06}. The error (in percent) of the prediction of $B_9$ with the
set $[2,3]$ is of about $0.035 \%$ and $0.54 \%$ in $D=2$ and $D=3$ respectively.

\begin{table}
\caption[t1]{Virial coefficients for a hard spheres in $2$ and $3$ 
dimensions given in Table I of \cite{Clis06} and predictions 
using PPT with different sets.}
\begin{tabular}{|c|c|c|c|c|c|}
\colrule
\multicolumn{1}{|c|}{} 
&\multicolumn{1}{c}{$B_9/B_2^8$}
&\multicolumn{1}{c|}{$B_{10}/B_2^9$}
&\multicolumn{1}{c}{$B_9/B_2^8$}
&\multicolumn{1}{c|}{$B_{10}/B_2^9$} \\
\multicolumn{1}{|c|}{} 
&\multicolumn{2}{|c|}{$D=2$}
&\multicolumn{2}{c|}{$D=3$} \\
\colrule
Ref.\cite{Clis06}&         $0.0362193$  & $0.0199537$ & $0.0013094$ & $0.0004035$\\
\colrule
$[0,5]$ & $0.03739998$ & $0.02496595$ &  $0.0023400$ & $0.0031580$ \\
$[1,4]$ & $0.03625994$ & $0.02008503$ &  $0.0013509$ & $0.0004884$ \\
$[2,3]$ & $0.0362321$  & $0.0199843$  &  $0.0013165$ & $0.0004198$ \\
$[3,2]$ & $0.0362551$  & $0.02006717$ &  $0.0013404$ & $0.0004664$ \\
$[4,1]$ & $0.0368599$  & $0.02258546$ &  $0.0017325$ & $0.0014442$ \\ 
$[5,0]$ & $0.1747048$  & $0.75984885$ &  $0.0222648$ & $0.0735264$ \\
\colrule
\end{tabular}
\label{table2}
\end{table}

In Table \ref{table3} we have compared our predictions for the virial coefficients 
$B_{11}$ through $B_{18}$ using PPT with the set $[3,4]$ with the predictions
made in \cite{Clis06}, finding very similar results for $D=2$ but rather different
results in the case $D=3$.

\begin{table}[H]
\centering
\caption{Predicted coefficients for approximants with 10 exact 
         coefficients for $D=2$ and $D=3$. Comparison between 
         the predictions of \cite{Clis06} and the 
         predictions obtained using the set $[3,4]$.}
\label{table3}
\vspace{2ex}
\scriptsize
\begin{tabular}{c|cccccccc}
\colrule
\\[-1.5ex]
  &\multicolumn{1}{c}{$B_{11}/B_2^{10}$} &\multicolumn{1}{c}{$B_{12}/B_2^{11}$} &\multicolumn{1}{c}{$B_{13}/B_2^{12}$} &\multicolumn{1}{c}{$B_{14}/B_2^{13}$} &\multicolumn{1}{c}{$B_{15}/B_2^{14}$} &\multicolumn{1}{c}{$B_{16}/B_2^{15}$} &\multicolumn{1}{c}{$B_{17}/B_2^{16}$} &\multicolumn{1}{c}{$B_{18}/B_2^{17}$}\\[0.7ex]
\colrule
\\[-1.5ex]
$D=2$  & & & & & & & \\
Ref.\cite{Clis06}  & $1.089 \times 10^{-2}$  & $5.90 \times 10^{-3}$ & $3.18 \times 10^{-3}$ & $1.70 \times 10^{-3}$ & $9.10 \times 10^{-4}$ 
                   & $4.84 \times 10^{-4}$   & $2.56 \times 10^{-4}$ & $1.36 \times 10^{-4}$ \\
$[3,4]$            & $1.0901 \times 10^{-2}$ & $5.9235 \times 10^{-3}$ & $3.2117 \times 10^{-3}$ & $1.7421 \times 10^{-3}$ &  $9.4698 \times 10^{-4}$ 
& $5.1638 \times 10^{-4}$ & $2.8247 \times 10^{-4}$  & $1.5492 \times 10^{-4}$ \\
\colrule
$D=3$  & & & & & & & \\
Ref.\cite{Clis06}  &$ 1.22 \times 10^{-4}$ &$ 3.64 \times 10^{-5}$ &$ 1.08 \times 10^{-5}$ &$ 3.2 \times 10^{-6}$ &$ 9.2 \times 10^{-7}$ &$ 2.6 \times 10^{-7}$ &&\\
$[3,4]$ & $1.1599 \times 10^{-4}$ & $2.2229 \times 10^{-5}$ & $-8.5616 \times 10^{-6}$ &  $-1.8088 \times 10^{-5}$ & 
$-2.0325 \times 10^{-5}$ & $-2.0112 \times 10^{-5}$ &  $-1.9136 \times 10^{-5}$ & $-1.7971 \times 10^{-5}$ \\
\colrule
\end{tabular}
\normalsize
\end{table}

Concluding, our method has several interesting features: it can describe multivalued functions, it can provide the
imaginary part of the observable corresponding to a metastable state, it can resum divergent series and select
the most appropriate asymptotic behavior of the solution among those available at a given order. Finally, it 
can also be used to make accurate predictions of yet unknown perturbative coefficients. Further applications
of this method have been considered in a lengthier and more detailed paper.


\begin{thebibliography}{}
\bibitem{BW01} C.M.Bender and E.J.Weniger, J. Math. Phys.{\bf 42}, 2167-2183 (2001)
\bibitem{JZJ} J.Zinn-Justin, Quantum Field Theory and Critical Phenomena, Clerendon Press-Oxford, New York (2002);
I.R.C.Buckley, A. Duncan and H.F.Jones, Phys.Rev.{\bf D} 47, 2554-2559 (1993);
C.M.Bender, A. Duncan and H.F.Jones, Phys.Rev.{\bf D} 49, 4219-4225
\bibitem{Nish01} Y. Nishiyama, J.Phys.{\bf A} 34, 11215-11223 (2001)
\bibitem{Amore06a} P.Amore, J.Phys.{\bf A} L349-L355  (2006)
\bibitem{Clis06} N.Clisby and B. McCoy, Jour. of Stat. Phys. {\bf 122} 15-57 (2006)  
\end{thebibliography}
\end{document}